\documentclass[preprint,eqsecnum,aps,showpacs]{revtex4}
\usepackage{dcolumn}
\usepackage{graphicx}
\usepackage{amsmath}
\usepackage{amssymb}
\usepackage{epsfig}
\usepackage{float}
\usepackage{latexsym}
\usepackage{rotating}

\begin{document}

\title{\small{Scalar field Pressure in Induced Gravity with Higgs Potential and Dark Matter}}

\vspace{1.0cm}
\author{Nils M. Bezares-Roder$^{1,3}${\footnote{Nils.Bezares@uni-ulm.de}}\, Hemwati
Nandan$^{2,3}${\footnote{hnandan.ctp@jmi.ac.in}} and Heinz
Dehnen$^3${\footnote{Heinz.Dehnen@uni-konstanz.de}}}
\vspace{0.5cm}
\affiliation{\small{$^1$Institut f\"{u}r Theoretische Physik, Universit\"at Ulm,$\,890\,69$ Ulm,
Germany\\
$^2$Centre for Theoretical Physics, Jamia Millia Islamia, New Delhi, 110\,025, India\\
$^3$Fachbereich Physik, Universit\"at Konstanz, $M 677, 784 \, 57$ Konstanz, Germany}}

\vspace{1.8cm}
\begin{abstract}
A model of induced gravity with a Higgs potential is investigated in detail in view of the pressure components related to the scalar-field excitations. The physical consequences emerging as an artifact due to the presence of these pressure terms are analysed in terms of the constraints parting from energy density, solar-relativistic effects and galactic dynamics along with the dark matter halos.\\

\vspace{0.2cm}

\end{abstract}

\vspace{1.0cm}
\maketitle
\section{Introduction}
\noindent Scalar-field models in cosmology and classical gravity have been quite successful in describing variety of important issues in modern cosmology (viz the problem of dark energy, dark matter, inflation, late time acceleration of the Universe, cosmic evolution etc.; see e.g. \cite{Sami, Sahni, Copeland, Sami1, Xu, Neupane, Das, Feng, Brun, Ratra, Padmanabhan, Kinney, Kozyrev08, Rod09, Bijay, Campo,Bibeka,Jae-Weon Lee1,Jae-Weon Lee2}. In particular, the use of a Higgs scalar field \cite{Higgs} in a Jordan--Brans--Dicke (JBD) model of gravity \cite{Brans} is of quite significant importance. It was first acquainted during the early nineties of the last century \cite{Dehnen1,Dehnen2,Dehnen3} by using the idea of the broken-symmetric theory of gravity \cite{Zee} along with a JBD model of gravity \cite{Brans}. Since then, it has seen various subsequent developments during the last few years \cite{Cervantes95a, Bezares07-DM, Cervantes07a, Bezares06H, nils-prom, Rod09, Bij94, Bezares09-RN}. Such a model is important especially because, as a result of spontaneous breakdown of symmetry, a Higgs potential as used in the model naturally leads to a cosmological function of anti-gravitational behaviour as well as because of the role of the Higgs field in the mass generation mechanism of the constituents, i.e. the basic building blocks (quarks, leptons and gauge bosons) of nature \cite{Peskin}.\\
Many efforts to find exact solutions of the field equations entailing scalar fields (with and without mass) in gravity have been made time and again \cite{Wyman, Schmoltzi, Jetzer, Hardell, Vir, Bijay}. Within GR, they lead to singularities and Black Holes related to Schwarzschild singularities which hide point--singularities from outside. In particular, whether naked singularities exist or not is still an open question, both within scalar--tensor theories and within GR. Actually, the \emph{Cosmic Censorship Conjecture} (CCC) \cite{Penrose98} prohibits such singularities, which are widely regarded as nonexistent. However, their appearance is theoretically predicted in several models of gravity \cite{Ori,Joshi1,JoshiB,Jhingan,Shapiro91,Zade}. Different classes of BH and wormhole solutions as well as the nonsingular spherically symmetric solution in such models have been constructed and analysed in different contexts \cite{Bronnikov,Sheykhi,Kozyrev,Nandi,moffat}. In case the CCC is violated, then the naked singularities might contribute to some new gravitational phenomena including their relationship to gamma-ray bursts (GRBs) \cite{Chakrabarti,Joshi2,Harko03}.\\
The appearance of singular and nonsingular Black Hole (BH) solutions in scalar--tensor theory models \cite{Haw72,Cam93,Bhadra,Bezares07-BH}, and their relation with the facets of the scalar field are of crucial interest \cite{Avelino,Lopez,Narayana}. Furthermore, we have noticed the breakdown of CCC (i.e. the appearance of naked singularities) in certain circumstances in our investigations of this induced gravity model with Higgs potential \cite{Bezares07-BH,Bezares09-RN}. Recently, we have studied the appearance of BHs as well as their consequences in a scalar--tensor theory with Higgs potential along with the amplitude (integration constant) of the vacuum scalar-field excitation in relation to the appearance of quintessential effects towards Reissner--Nordstr\"{o}m-like BH solutions \cite{Rei16} and flattening of the galaxy rotation curves \cite{Bezares07-DM,Bezares09-RN}. Further, the above mentioned integration constant indicates a relation to effective mass as measured mass which differs from bare (luminous) mass derived only from the usual densities \cite{Bezares09-RN}. This  integration constant of the scalar-field excitation is in fact related to the energy--stress tensor and hence to the densities and pressures. It therefore plays a crucial role in the dynamics of the present model of induced gravity. On the one hand, the energy density and pressure appear within linear solutions such that the finite pressure terms contributing from the scalar field account to the dynamical mass. These terms are regarded to assure the equivalence with Newtonian and standard results of GR having consistency within a Parameterised-post-Newtonian (PPN) framework \cite{Bezares09-RN}.\\
A dynamical mass parameter appears for linearised gravity, whereas an effective mass appears for higher-order solutions. Their interrelation as pressure containing terms is of special relevance as pressure screens bare mass from density with dynamical pressure terms which are related to the scalar field. On the other hand, for galactic dynamics, a Dark Matter profile has been derived for flat rotation curves \cite{Cervantes07a,nils-prom}. Within this framework,
density may be decomposed in a usual, mainly baryonic component and a further scalar-field density term which is directly related to a pressure term \cite{nils-prom}. It is of crucial interest to constrain such terms as well as to clarify their nature and relevance for galactic dynamics as well as for solar-relativistic effects, i.e. for solar ranges and higher-order solutions. In both cases, scalar-field pressures, i.e. pressure terms of the perfect fluid which are related to the appearance of the scalar-field excitation, are of special interest.\\
In the present paper, we address the above mentioned issues and point out a universal character of scalar-field pressure to understand the dark-matter problem. The paper is organised as follows: we first recall the field equations and the integration constant of scalar-field excitations corresponding to the model used for our investigations in Section I. The field equations are then identified with Maxwell-like equations of gravity with an energy density in Section II. Utilising these equations, the physical consequences of the scalar-field pressure are presented for linear solutions and perihelion advance in the first and second parts of Section III respectively. Further, in the last part of Section III, we present the dark-matter profile as well as the constraints on pressure terms coming from dark-matter phenomenology within this model. The non-Newtonian properties of the density are also discussed for different length scales. Finally, we conclude our findings in Section IV with some interesting issues which need further investigations.

\section{Field equations and scalar field pressure}
\noindent Let us consider the following action \cite{Dehnen3,Bezares07-BH},
\begin{equation}
    {\cal S}= \int d^4 x\, \sqrt{-g}\, \left[\frac{\gamma}{16\pi}\,
    \phi^\dagger \phi \, {R} \,+ \, \frac{1}{2} \, \phi^\dagger_{;\mu} \,
    \phi^{; \mu} - V(\phi) + {\cal L}_M (\psi, {\cal A}_\mu, \phi)
    \right], \label{action}
\end{equation}
where $\gamma$ is a dimensionless constant and ${\cal L}_M (\psi, {\cal A}_\mu, \phi)$ is the Lagrangian corresponding to matter, i.e. the fermionic $(\psi)$ and massless bosonic (${\cal A}_\mu$) fields. Here, semicolon is used for a covariant derivative. The self-interacting Higgs potential $V(\phi)$ characterises the varying cosmological and gravitational ``constants''. The potential is normalised such that it realise Zee's assumption $V(\phi = v) =0$ for the ground-state value of $\phi$ \cite{Zee} in absence of the cosmological constant. In the presence of a cosmological term $\Lambda_0$, it possess the following form,
\begin{equation}
    V(\phi)= \frac{\mu^2}{2} \, \phi \, \phi^\dagger + \frac{\lambda}{4!}
    \, (\phi^\dagger \phi)^2+ \bar{V}+ V_0, \label{H}
\end{equation}
where $\mu^2 < 0$ and $\lambda > 0$ are real-valued constants and $\bar{V} = 3 \,\mu^4 / {(2 \lambda)}$ \cite{Dehnen1}. Further, we have $V_0= -3\gamma \mu^2 \Lambda_0/(4\pi \lambda)$ entailing a cosmological constant $\Lambda_0$ \cite{nils-prom,vd2009}. The Higgs field in the spontaneously broken phase of symmetry leads to the square of the ground-state value as $v^2= \phi_0\,\phi_0^{\dagger}= -\, {6\mu^2}/{\lambda}$ and can further be resolved as $\phi_0=v N$ (where $N$ is a constant which satisfy $N^\dagger N=1$) with the introduction of the unitary gauge \cite{Dehnen2}. The general Higgs field $\phi$ may then be redefined through a real-valued excitation ($\xi$) of the Higgs scalar field as $ \phi=v \,(1+ \xi)^{1/2} \,N $. One may notice that in absence of the scalar-field excitations $\xi$, $V_0$ leads to the usual cosmological constant $\Lambda_0$ of $\Lambda CDM$.\\
The general form of the matter Lagrangian which will be used has the following form from the prespectives of elementary particles \cite{Cervantes95a,nils-prom},
\begin{equation}
 {\cal L}_M= - \frac{1}{16 \pi}\, {\cal F}_{\mu \nu} {\cal F}^{\mu \nu} + \frac{i}{2} \, \,\bar{\psi}\, \gamma^\mu_{_{L,R}} \, \psi_{;\mu}+ h.c.- (1- \hat{q})\, k\, \bar{\psi}_{_{R}} \,{\varphi}^\dagger \hat{x}\, \psi_{_{L}}+ h.c. \label{matterlag}
\end{equation}
This matter Lagrangian is related to the energy-stress tensor and the closure of the system. Within the notions of elementary-particle physics \cite{Dehnen1}, it entails the left $(L)$ and right handed $(R)$ fermionic ($\psi$) states with Dirac conjugate $\bar\psi$, Dirac matrices $\gamma^\mu$ and field-strength tensor $F_{\mu \nu}$ in matrix representation ${\cal F}_{\mu \nu} = {\cal A}_{\nu,\mu} - {\cal A}_{\mu,\nu} + ig [{\cal A}_\mu,{\cal A}_\nu]$, where ${\cal A}_\mu$ denote the massless bosonic gauge fields in matrix representation. Here $\hat x$ is the Yukawa coupling operator. We have further introduced a parameter $\hat q$ to give the fermionic coupling with the scalar field \cite{Dehnen1}. We consider $\varphi=\phi$ in (\ref{matterlag}), which means that the same scalar field couples with the Ricci scalar and matter for the case $\hat q \neq 1$. If both the non-minimally coupled scalar Higgs field and the Yukawa-coupled Higgs field which leads to mass generation of fundamental matter are the same and the cosmological constant $\Lambda_0$ is zero, the source of the scalar-field excitation will vanish after symmetry breaking, as it cancels out exactly due to the appearing energy-stress $T=\sqrt{1+\xi}\,\bar{\psi}\,\hat{m} \, \psi$, with the fermionic mass matrix $\hat{m}$ related to the Yukawa coupling (see \cite{Dehnen3} for more details).\\
In the general context of gauge fields, to the energy--momentum tensor $T_{\mu \nu}$ further gauge-field terms ${\cal A}_\mu$ multiplied to the mass-square matrix ${\cal M}^2$ are to be added onto an effective tensor $\hat{T}_{\mu \nu}$. However, the explicit form of gauge fields related to the covariant derivative depend on the explicit model. Especially within the SM, since left-handed states underlie different gauge groups, they differ for left- and right-handed fermionic fields (but also depend on the type of fermion). However, there are no derivatives in the Yukawa term and, thus, both-handed states acquire equal mass. There appear only a coupling constant and the Yukawa matrix. The latter leads to different masses of both leptons and quarks.\\
The covariant derivative in equation (\ref{matterlag}) is a gauge and gravity covariant derivative entailing gauge fields as well as Christoffel and Ricci coefficients of spacetime curvature as connections. Suppressing gauge fields, the scalar-field equation (i.e. the Klein--Gordon equation) is obtained as given below,
\begin{align}
    \xi^{,\mu}\,_{\,;\,\mu}+ \frac{\xi}{L^2}\, =\,
    \left(1+\frac{4\pi}{3\gamma}\right)^{-1}\cdot\frac{8\pi G}{3}\hat{q}T +\frac{4}{3}\,\left(1+
    \frac{4\pi}{3\gamma}\right)^{-1}\Lambda_0.\label{thp:Higgs01}
\end{align}
The Higgs field in the broken phase of symmetry in this model possesses a finite range $L$ (i.e. Compton wavelength) which in the natural (geometric) system of units is inverse of our Higgs field mass \cite{Dehnen3,Bezares07-DM,Bezares09-RN} as given below,
\begin{equation}
    L =\left[ { \frac{1 + \frac {4 \pi }{3 \gamma } } {16 \pi
    G ( \mu^4 / \lambda ) } } \right]^{1/2}\left(\frac{\hbar}{c}\right)\,,
\end{equation}
with the gravitational constant $G=1/({\gamma v^2})$ where $\gamma \gg 1$ is defined as the square of the ratio of the Planck $(M_P)$ and gauge-boson $(M_A)$ masses \cite{Bezares07-BH, Dehnen2}. However, the effective gravitational coupling is given in terms of the scalar-field excitations. It is defined as $\tilde {G} = {G}/({1+ \xi})$.\\
We have {(i)} $\hat{q}=0$ for a coupling of the scalar field to the fermionic Lagrangian \cite{Dehnen3,Cervantes95a}, {(ii)} for no such coupling ($\hat{q}=1$) (i.e. the source is weak) (cf. \cite{Bezares07-DM, Cervantes95a}). For the physical properties of the particles related to this Higgs field, the case (i) means that the particles, which are responsible for mass of elementary particles, decouple and interact only gravitationally for $\Lambda_0=0$. Such particles may not be generated through high-energy collision experiments as suspected to observe in the LHC experiments. They may, however, be related to a cosmological constant $\Lambda_0$ as source. On the other hand, the case (ii) means new particles which interact with other particles indeed, however weakly. If they are as massive as indicated in \cite{Bezares07-DM, Cervantes07, nils-prom}, then they hardly decay in less massive particles and are rather stable \cite{Bij94}. Furthermore, with the low masses, they still lie below the accuracy range of \emph{fifth- force} experiments \cite{Adelberger} and to-date do not indicate a breakdown of the equivalence principle.\\
However, the energy-stress tensor satisfies the following equation law,
\begin{align}
  T_\mu\,^\nu\,_{;\nu}= (1- \hat{q}) \frac{1}{2}\xi_{,\mu}(1+ \xi)^{-1} T.\label{erhalt}
\end{align}
In the case that $\phi$ does not couple to the fermionic state in ${\cal L}_M\sqrt{-g}$ (i.e. $\hat{q}=1$), then equation (\ref{erhalt}) will not possess a source, and for the SM ($\hat{q}=0$), above equation means the production of fermionic mass through this Higgs field. This leads to a breaking of the conservation law through a new \emph{Higgs force}.\\
\noindent The Einstein equations corresponding to the action (\ref{action}) acquire the following general form,
\begin{alignat}{1}
    R_{\mu \nu}- \frac{1}{2}R g_{\mu \nu}+ \frac{3}{4L^2}\, \frac{\xi^2}{1+\xi}\, g_{\mu \nu}+ & \frac{\Lambda_0}{1+\xi} \, g_{\mu \nu}= -\frac{8\pi \tilde{G}}{c^4} \, T_{\mu \nu}- \frac{1}{1+\xi} \, \left[\xi_{,\mu; \, \nu}- \xi^{,\lambda}\,_{;\lambda} \, g_{\mu \nu}\right]-\label{Einstein}\\
    &- \frac{\pi}{\gamma} \, \frac{1}{(1+\xi)^2} \, \left[2\xi_{,\mu} \, \xi_{,\nu}- \xi_{,\lambda} \,\xi^{,\lambda} \, g_{\mu \nu}\right].\nonumber
\end{alignat}
Now, using the scalar-field equation (\ref{thp:Higgs01}), equation (\ref{Einstein}) may further be transformed into the form given below,
\begin{alignat}{1}
    R^\sigma\,_{\nu} \, u_\sigma-& \frac{1}{2}R \, u_\nu+ \frac{1}{L^2} \, \xi \left(1+\xi\right)^{-1}\left[1+\frac{3}{4} \, \xi\right]u_\nu+ \frac{\Lambda_0}{1+\xi}\left[1- \frac{4}{3}\, \left(1+ \frac{3\pi}{4\gamma}\right)^{-1}\right]u_\nu\label{Einsteinu2}\nonumber\\
   \hspace{-2.5cm} &= -\tilde{\kappa} j_\nu +\left(1+ \frac{4\pi}{3\gamma}\right)^{-1} \frac{T}{3}\,\hat{q}\, u_\nu- \frac{1}{1+ \xi} \, \xi^{,\sigma}\,_{;\nu} \, u_\sigma- \frac{\pi}{\gamma} \, \frac{1}{1+ \xi}\left[2 \, \xi^{,\sigma}\xi_{,\nu} \, u_\sigma- \xi_{\lambda} \, \xi^{,\lambda}\, u_\nu\right],
\end{alignat}
with $\tilde{\kappa}=\kappa_0/(1+\xi)$ and $\kappa_0=8\pi G/c^4$. Here, we have defined the current by means of the energy--momentum density of matter measured by the observer in the following form,
\begin{align}
    j_\mu\equiv T^\sigma\,_\mu \, u_\sigma.\label{j}
\end{align}
This current may be related to an equation analogue to Maxwell's one of electrodynamics. Actually, with the observer field $u_\nu$, if a (to mass parametrised) non-gravitational force is given by $K_\mu=u_{\mu;\lambda}u^\lambda$, the existence of an equilibrium between that force and inertial forces will maintain the mass in a geodesic trajectory such that $E_\mu+ K_\mu=0$ is valid for an inertial force $E_\mu$. The inertial force $E_\mu$ is then given as follows,
\begin{align}
    E_\mu= -u_{\mu;\sigma}u^\sigma= (u_{\sigma; \mu}- u_{\mu;\sigma})u^\sigma= \tilde{F}_{\mu \sigma} u^\sigma\,,\label{fieldst}
\end{align}
where $\tilde{F}_{\mu \nu}= u_{\nu;\mu}- u_{\mu;\nu}$ is a field-strength tensor (having an analogue structure to electrodynamics) and $u^\mu$ a gauge variable \cite{Dehnen64a}. For the field-strength tensor a relation analogous to the electrodynamics is therefore evident as given below,
\begin{align}
    \tilde{F}_{(\lambda \mu, \nu)}\equiv \,&\tilde{F}_{\mu \nu, \lambda}+ \tilde{F}_{\lambda \mu,\nu}+ \tilde{F}_{\nu \lambda,\mu}=\, 0.\label{Maxweq}
\end{align}
Further, for equation (\ref{Einsteinu2}), the following relation is thus valid \cite{Dehnen64a},
\begin{align}
    \tilde{F}_\mu\,^\lambda\,_{;\lambda}=2\tilde{\kappa}\left(j_\mu+ s_\mu\right),
\end{align}
where $s_\mu$ is the energy--momentum density of the gravitational field. With these considerations, $s_\mu$ has the following form,
\begin{alignat}{1}
    s_\nu=& -\frac{1}{2\tilde{\kappa}}\left\{\left[\tilde{\kappa} \left(1-\frac{\hat{q}}{3}+ \frac{4\pi}{3\gamma}\right)T- \frac{1}{L^2}\left(\frac{1+\frac{3}{2}\xi}{1+\xi}\right)\xi- \frac{2}{3}\left(1+ 4\frac{\pi}{\gamma}\right)\frac{\Lambda_0}{1+ \xi}\right]u_\nu+ Q_\nu\,^{\sigma}\,_{;\sigma}\right\}+\nonumber \\
    &+ \frac{1}{\kappa_0}\left[\xi^{,\sigma}\,_{;\nu} + \frac{\pi}{\gamma}(1+ \xi)^{-1} \left(2\xi^{,\sigma} \xi_{,\nu}- \xi_{,\lambda}\xi^{,\lambda}\delta^\sigma\,_\nu \right)\right]u_\sigma,
\end{alignat}
where the $Q_\mu\,^\nu$ is a second-rank tensor which is related to the Ricci tensor. In fact, the Ricci tensor can be constructed in terms of the antisymmetric field-strength tensor $\tilde{F}_{\mu \nu}$ and the symmetric tensor $Q_{\mu \nu}$ (for details see {\bf Appendix I}).\\
The field strength $\tilde{F}_{\mu \nu}$ has two sources: (i) 4-currents $j_\nu$ as energy--momentum density of matter, and (ii) $s_\nu$ as energy--momentum density of the gravitational field \cite{Dehnen64a}. Consequently, momentum conservation is valid with the following equation,
\begin{align}
  (j^\mu+ s^\mu)_{;\mu}= 0.\nonumber
\end{align}
The energy density measured by a static observer is $s=s_\mu u^\mu$ which for the case $\gamma\gg 1$ is derived as given below,
\begin{alignat}{1}
  s=& \frac{2}{\tilde{\kappa}}u_\mu\,^{;\sigma} u^\mu\,_{;\sigma}+ \frac{1}{\kappa_0}\xi^{,\mu}\,_{;\mu}+ \nonumber\\
  &+ \frac{1}{\tilde{\kappa}}\left\{u^\sigma\,_{;\mu ;\sigma} u^\mu- \frac{\tilde{\kappa}}{2}\left(1- \frac{\hat{q}}{3}\right)T+ \frac{1}{2L^2} \left(\frac{1+ \frac{3}{2}\xi}{1+ \xi}\right)\xi+ \frac{1}{3}\frac{\Lambda_0}{1+ \xi}\right\}.\label{s17}
\end{alignat}
For weak dynamical behaviour, the metric is nearly constant, and for static fields (i.e. $u^\lambda\,_{;\lambda}= (\sqrt{-g} u^\lambda)_{,\lambda}/\sqrt{-g}=0$), the Ricci identities lead to
\begin{align}
  R_{\mu \nu}u^\mu u^\nu= -u^\sigma \,_{;\mu ;\sigma} u^\mu .\label{ru}
\end{align}
Further, the Ricci scalar $R$ may be derived by taking the trace of equation (\ref{Einstein}) in the following form,
\begin{align}
  R= \frac{3}{L^2}\xi + 8\pi \tilde{G} T \left(1- \frac{\hat{q}}{1+ \frac{4\pi}{3\gamma}}\right)- \frac{2}{\gamma}\frac{\xi_{,\lambda}\xi^{,\lambda}}{(1+\xi)^2}+ 4\frac{\Lambda_0}{1+\xi}\left(1- \frac{1}{1+ \frac{4\pi}{3\gamma}}\right).
\end{align}
Now the equation (\ref{ru}) reads as follows,
\begin{align}
  R_{\mu \nu}u^\mu u^\nu= -\frac{1}{2L^2}\left(\frac{1+ \frac{3}{2}\xi}{1+ \xi}\right)\xi- \frac{1}{3}\frac{\Lambda_0}{1+ \xi}+ \tilde{\kappa} T_{\mu \nu} u^\mu u^\nu- \frac{\tilde{\kappa}}{2}\left(1- \frac{\hat{q}}{3}\right)T+ \frac{1}{1+ \xi}\,{\xi^{,\mu}\,_{;\mu}}.\label{sal}
\end{align}
Using the scalar field equation (\ref{thp:Higgs01}) in the equation (\ref{s17}) with $u^\mu$ as timelike, we finally arrive to the following expression for the energy density of the gravitational field,
\begin{align}
    s= \frac{2}{\tilde{\kappa}} u_\mu\,^{;\sigma}u^\mu\,_{;\sigma}+ T_{\mu \nu} u^\mu u^\nu- \left(1- \hat{q}\right)T- \frac{2}{\kappa_0 L^2}\xi+ \frac{8}{3}\frac{\Lambda_0}{\kappa_0}\,.\label{scov}
\end{align}
Within the present model, it is feasible to assume the vanishing parameters $\lambda$ and $\mu$ of the Higgs potential (\ref{H}) without a vanishing ground state value as a consequence \cite{Bezares07-BH,Bezares09-RN}. Further, for small $\lambda$ \cite{Bij94}, the mass $M$ of the Higgs field would become very small, resulting in a contribution to the gravitational force with a range proportional to $1/\sqrt{\lambda}$ with the Higgs particle behaving as the cosmon of Quintessence.\\
For $L\rightarrow \infty$ (or low scalar field excitations), together with $\hat{q}=0$ and $\Lambda_0=0$, equation (\ref{scov}) gives the usual energy density. For $\hat{q}=1$, however, there is no term proportional to $T$ such that $\kappa_0$ is to be rescaled to $\kappa_N$ (viz \cite{Dehnen2, Cervantes07a, Bezares09-RN}).\\
If we consider the energy--momentum of an ideal fluid then there is $T=T_{\mu \nu}u^\mu u^\nu- 3p$ along with $T=\epsilon- 3p$ with energy densities $\epsilon= \varrho c^2$ and pressures $p$ and the 4-velocity $u_\mu$. With these considerations, energy density of gravitation as in (\ref{scov}) now leads to the form as given below,
\begin{align}
    s= \frac{2}{\tilde{\kappa}}u_\mu\,^{;\sigma} u^\mu\,_{;\sigma}+ \hat{q}\epsilon+ 3(1- \hat{q})p- \frac{2}{\kappa_0 L^2}\xi+ \frac{8}{3}\Lambda_0\,.\label{s}
\end{align}
With the Cartesian coordinates for central symmetry, for an observer which is static to matter, there is a 4-velocity in linear approximation as follows,
\begin{align}
    u_\mu=\left(1- \frac{\nu}{2},0,0,0\right).
\end{align}
With $\nu c^2/2=\Phi$ (the gravitational potential), we have
\begin{align}
  u_\mu^{;\sigma} u^\mu\,_{;\sigma} c^4= -(u^0\,_{,1})^2c^4= -(\text{grad}\, \Phi)^2.
\end{align}
The equation (\ref{s}) therefore reads in the static case as follows,
\begin{align}
    s=- \frac{2/c^4}{\tilde{\kappa}}(\text{grad}\, \Phi)^2+ \hat{q}\epsilon+3(1- \hat{q})p - \frac{2}{\kappa_0 L^2}\xi+ \frac{8}{3\, \kappa_0} \Lambda_0\,.\label{sss}
\end{align}
The linear Einstein equation for $\nabla^2 \Phi$ (see {\bf Appendix II} for details) together with the scalar field equation lead to a Poisson equation \cite{Bezares09-RN} which may be written as follows in case it is not linearised for $\xi$,
\begin{align}
  \nabla^2 \Phi(1+\xi)+ \frac{c^2}{2}\left(1+ \frac{3}{2}\xi\right) \nabla^2 \xi= \frac{\kappa_0}{3}\left[3\epsilon- \frac{3}{2}(\epsilon- 3p)- \frac{3}{4}\xi \hat{q}(\epsilon- 3p)\right].\label{Poisson}
\end{align}
For a perfect fluid and using the pressure-comprising Poisson equation (\ref{Poisson}), we may write the following for weak fields and non-dominant $\xi$ excitations,
\begin{align}
    \text{grad}\, p= -\frac{2}{\kappa_0}\text{grad}\, \Phi \left(\frac{\nabla^2 \Phi}{c^4}- \frac{3}{2}p\,c^2\right).\label{gradp}
\end{align}
This problem is thus analogous to the one of GR plus a pressure term. Now, under the assumption $L\rightarrow \infty$, using the Gauss theorem several times and taking into account that the pressure $p$ is supposed to vanish at the surface of matter distribution, equation (\ref{gradp}) leads to a relation between the gravitational potential and pressure terms $p$ as below,
\begin{align}
    3c^4\int p dV=\frac{1}{\kappa_0}(1+ 3w)^{-1} \int (\text{grad}\, \Phi)^2 dV,\label{p1w}
\end{align}
where $p=w \epsilon$ with an equation-of-state (EOS) parameter $w$ \cite{Dehnen64b}. The equation (\ref{p1w}) provides a relationship between the Newtonian gravitational pressure in matter and the gravitational field strength. The equation (\ref{p1w}) is related to the gravitational energy--momentum density by equation (\ref{sss}). Actually, for the field energy, for weak $\xi$ fields with $L\rightarrow \infty$ and a vanishing cosmological constant $\Lambda_0$, there is
\begin{align}
    S= \int sdV= -\frac{2/c^4}{\kappa_0} \int (\text{grad}\, \Phi)^2dV+ \hat{q}\int \epsilon\, dV + 3(1- \hat{q}) \int p dV,\label{p1wa}
\end{align}
with an energy term as defined below,
\begin{align}
    c^4\int \epsilon dV=\frac{1}{3\kappa_0} \frac{1}{w (1+ 3w)} \int (\text{grad}\, \Phi)^2 dV.\label{p1ww}
\end{align}
Using equation (\ref{p1ww}) in (\ref{p1wa}), the field energy (\ref{p1wa}) then yields
\begin{align}
  S = -\frac{c^4}{\kappa_0} \left[2- \frac{\hat{q}}{3(1+ 3w)}- \frac{(1- \hat{q})}{(1+ 3w)}\right]\int \left(\text{grad}\, \Phi\right)^2dV.\label{s1}
\end{align}
The solution within GR reads $S=-(c^4/\kappa_N) \int (\text{grad}\, \Phi)^2 dV$. It gives the same as the potential energy of a body within Newton's gravitational theory, and such is necessary to avoid conflicts with elementary mechanics \cite{Dehnen64a, Dehnen64b}. The equation (\ref{sss}) gives GR's solution for $\hat{q}=0$ and $w=0$ with $\kappa_0=\kappa_N$ indeed. For $\hat{q}=1$, on the other hand, after rescaling with $\kappa_N= 4\kappa_0/3$,
equality between the usual gravitational energy of GR and induced-gravity's gravitational energy (as of interpretation given here) leads to a constraint of $w\approx 0.17 \approx 1/6$. This pressure value appears necessary for consistency with phenomenology and for non-vanishing scalar fields. Hence, it would appear in some specific contexts, and variations from it would
lead to measurable consequences especially for large-scale dynamics. In general, this may respond to nature of Quintessence fields whose fluctuations may behave similar to a relativistic fluid \cite{Wetterich02}. Further, given that models of Quintessence usually predict composition-dependent gravity such as long-range forces mediated by the fields \cite{Wetterich03}, measurable consequences would appear for large distances indeed. The scalar fields may act similar to a cosmological constant (cf. \cite{Doran02}) or as related to the
halo mass of galaxies and hence to Dark Matter phenomenology \cite{Wetterich01}.

\section{Consequences of scalar field pressures}

\subsection{Linear Solutions}
\noindent For vanishing scalar-field excitations, equation (\ref{Poisson}) reduces to that of the usual GR. Consequently, the scalar field acts as a further gravitational interaction which at low scales is of Newtonian form. Further, it leads to an effective (measured) mass which possesses
scalar-field contributions via pressure $p$ \cite{Bezares09-RN}. Vacuum (linear) solutions $r\gg R_1$ with radii $R_1$ of the gravitational source (i.e. the massive objects) are of the same form as within GR, with additional terms of the integration constants entailing pressure of the gravitational object. The potentials in the components of metric (\ref{element}) are then
given as follows \cite{Bezares09-RN},
\begin{align}
   \nu= - \frac{r_{dyn}}{r},\quad \lambda=h\frac{r_{dyn}}{r},\label{pots}
\end{align}
The potentials given in equations (\ref{pots}) in principle show the usual GR behaviour for a dynamical Schwarzschild radius $r_{dyn}$, however, with a deviation expressed by the function $h(w)\equiv h$. The function $h$ is dependent on $r/L$ and it shows that, independently on $L$, for $\hat{q}=1$, there is $e^\nu\neq e^{-\lambda}$ unless there are finite values of the pressure $p$ which enter vacuum dynamics through the continuity conditions at $r=R_1$. This is a consequence of the scalar-field equation with an integration constant which depends on the trace of the energy--stress tensor \cite{Bezares07-BH}. Given the large scales $L$ compared to distance $r$ as assumed here, we obtain the asymptotic form of $h(w)$ as
\begin{align}
  h(w)=\frac{1+ 8w}{2+ 3w}.\label{h}
\end{align}
At distances $r$ of the order of the length scale $L$, i.e. for galactic spirals (as expected in \cite{Cervantes07,Bezares07-DM}), however, the value given by equation (\ref{h}) should be higher, yet, of the same order of magnitude.\\
Further, in equation (\ref{pots}) we have defined a dynamical Schwarzschild radius
\begin{align}
   r_{dyn}= \frac{2M_{dyn}G_N}{c^2} \label{sradius}
\end{align}
which is related to a dynamical mass as follows for the asymptotic limit $L\gg r$,
\begin{align}
   M_{dyn}= \left(1+ \frac{3}{2}w\right) M_1. \label{dmass}
\end{align}
Further, $M_1$ is the Schwarzschild mass obtained for vanishing pressure terms and vanishing scalar-field masses ($1/L=0$). Hence, $M_1$ is the Schwarzschild mass for the GR case. It appears as an integration constant of the gravitational potential (i.e. $\Phi\sim \nu$) of vacuum within GR and entails properties of the gravitational body by means of the integral of density
throughout the volume of the massive sphere. It is related to continuity conditions at the surface and, given the EOS as $p=w \epsilon$, the EOS parameter $w$ appears as a further mass term in $M_{dyn}$, which is related to pressure terms.\\
For weak-field approximation, consistency with a PPN framework is given for $h=1$ so that dynamical masses merge almost exactly. Both the equations (\ref{dmass}) and (\ref{sradius}) further simplify to the parameters as mentioned in \cite{Bezares09-RN} for $L\rightarrow \infty$. Further, for $L\rightarrow \infty$, a value of $w=1/5$ (viz the energy constraint from
Section II) in fact leads to $h= 1$. Furthermore, according to \cite{Bezares09-RN}, $h(w)r_{dyn}$ for high pressures is better given by the effective radius $\tilde{r}_S$ which indeed leads to the usual Schwarzschild radius for $w=1/6$ (see Subsection B below). Hence, energy arguments along with the effective-mass relation in view of exact solutions lead to non vanishing pressure terms related to $\xi$ which, as discussed in \cite{Bezares09-RN}, give an equation-of-state parameter of $1/6<w<1/5$ (see Subsection B and \cite{Bezares09-RN} for further details). This value range further leads to the usual GR relation between $\lambda$ and $\nu$. A detailed discussion about the evolution of the metric components and their relation to the usual solution may be found in \cite{Bezares09-RN}.\\
\begin{figure}[h] \centering
    \includegraphics*[width=14.5cm]{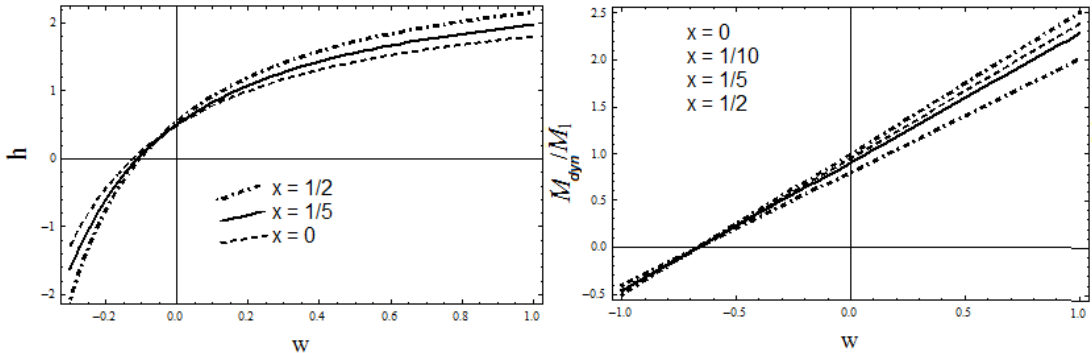}
    \caption{Evolution of the parameter $h$ and the dynamical mass coefficient $M_{dyn}/M$ for different EOS parameters $w$ and $x=r/L$.}
    \label{fig1}
\end{figure}
Fig. \ref{fig1} shows the basic behaviour of the correction term $h$ and $M_{dyn}/M_1$ in dependence of the EOS parameter $w$. For the sake of completeness, we plot these parameters for different $r/L$ relations. It is shown that $h$ possesses a larger value for higher pressures. At the same time, dynamical mass grows with pressure in respect to the bare luminous mass which comes solely from density. For $w=1$ and $L\rightarrow \infty$, for instance, the dynamical mass reads 2.5 times $M_1$ with $2hM_{dyn}\approx 9\cdot M_1$. Be reminded (cf. \cite{Bezares09-RN}) that $hM_{dyn}$ mainly gives the effective Schwarzschild mass $\tilde{M}_S$ which appears in the context of higher-order calculations. Further, it is possible to achieve flat rotation curves for polytropic density distributions with and without very massive galactic centres \cite{Bezares07-DM}. High pressure terms have a flattening behaviour for homogeneous density distributions \cite{Bezares09-RN}.\\
Effective, dynamical mass at long ranges does not have to equal the luminous mass which comes from density. Therefore, such a scenario would be important in the context of the Dark Matter problem and dark-matter phenomenology \cite{Kroupa}. Here, we analyse this possibility within the context of universal halo profiles (cf. \cite{NFW}) for flat rotation curves below in the subsection \ref{frc}.\\

\subsection{Perihelion Advance}
\noindent  In this part, we neglect scalar-field mass on the basis of discussions as of \cite{Bezares07-BH,Bezares07-DM,Cervantes07a,Bezares09-RN} and above. However, unlike in the previous part of this section, solar-relativistic effects need of higher-order corrections of the
time-coordinate related to the metric component. Hence, we will consider the solution already derived in \cite{Bezares09-RN} for further analysis.\\
In the vanishing limit of Higgs-like particle mass i.e. large distances, the scalar field equation (\ref{thp:Higgs01}) has the following form,
\begin{align}
  \xi^{,\mu}\,_{;\mu}= \frac{1}{\sqrt{-g}}\left(\sqrt{-g}\,\xi^{,\mu}\right)_{,\mu}= \hat{q}
\frac{8\pi}{3}G\,T. \label{mumu}
\end{align}
We will consider the determinant $g$ of the metric with the spacetime having spherical symmetry as given by the following line element,
\begin{align}
    ds^2=e^{\nu (r)} \,dt^2- e^{\lambda(r)} dr^2- r^2 \,(d \vartheta^2 + \sin^2 \vartheta \, d\varphi^2)\,. \label{element}
\end{align}
Using the line element (\ref{element}) and solving equation (\ref{mumu}) leads to the first integral of the scalar-field excitation as $\xi'=(A/r^2)\, e^{ (\lambda- \nu)/2}$ for vacuum \cite{Bezares07-BH,Bezares09-RN}, and the dimensionless integration constant $A$ appearing therein is defined as follows in the asymptotic limit $r\rightarrow \infty$ along with $L\rightarrow \infty$ \cite{Bezares09-RN},
\begin{align}
    A= \,-\,\frac{2}{3} \, \frac{G}{c^2}\int T \,\sqrt{-g} \, d^3
    x.\label{a}
\end{align}
Further, for the metric component $e^\nu$, there appears an integration constant $B$ which is related to the dynamical Schwarzschild radius with $e^\nu=1- B/r$.\\
For $A$ and $B$, there further appears an effective Schwarzschild radius
\begin{equation}
    \tilde{r}_S=2A+ r_{dyn}= \frac{2M_1G_N}{c^2}\left(\frac{1}{2}+ 3w\right)\approx h(w) r_{dyn}
\end{equation}
which is related to the dynamical mass in (\ref{pots}), and a squared generalised charge-parameter (a Reissner--Nordstr\"{o}m-like term) radius,
\begin{equation}
    r_Q =|\tilde{Q}^2|= \frac{|A\tilde{r}_S|}{2}.
\end{equation}
Further, $M_1$ represents the bare (luminous) mass, and the mass coefficient
$r_{dyn}/\tilde{r}_S$ is unlike one for $A\neq 0$.\\
The Reissner--Nordstr\"{o}m-like solution as derived in \cite{Bezares09-RN} reads as given below,
\begin{equation}
  e^\nu= \left[1- \frac{\tilde{r}_S}{r}\right]^{r_{dyn}/\tilde{r}_S};\,\, \, \quad e^\lambda= \left[1- \frac{\tilde{r}_S}{r}+ \frac{r_Q^2}{r^2}\right]^{-1},\label{potsrn}
\end{equation}
This solution may be used for analysis of the influence of the contributions from $A$ entailing further density and pressure terms from the inner field especially in terms of the linear solution within solar-relativistic effects.\\
Geodesics are the applicable trajectories for the theory, and for a well considered system in order to obtain the curves along a plane, for $\tilde{r}_S\ll r$, equation (\ref{potsrn}) leads to a Lagrange function of geodesic motion of the following form for the case $\hat{q}=1$,
\begin{align}
    \int {\cal L}d^3x = \frac{m}{2}\left[\left(1- \frac{\tilde{r}_S}{r}\right)^{r_{dyn}/\tilde{r}_S} \left(\frac{dx^0}{d\tau}\right)^2- \left(1+ \frac{\tilde{r}_S}{r}\right)\left(\frac{dr}{d\tau}\right)^2- r^2\left(\frac{d\varphi}{d\tau}\right)^2\right],\label{Lagfun}
\end{align}
with the eigentime $\tau$. Here, the cyclic coordinate $\varphi$ leads to a constant conjugate momentum as given below,
\begin{align}
    J=mr^2\frac{d\varphi}{d\tau}=mC_b^2= constant.\label{cb}
\end{align}
Further, cyclic coordinate $x^0=ct$ leads to
\begin{align}
    -m\left(1- \frac{\tilde{r}_S}{r}\right)^{r_{dyn}/\tilde{r}_S}\frac{dct}{d\tau}= mC_a= constant\label{ca}
\end{align}
which is valid for a parametrised energy term. Consequently, from equation (\ref{Lagfun}), we obtain the following relation,
\begin{align}
    \left(1+ \frac{\tilde{r}_S}{r}\right) \left(\frac{dr}{d\tau}\right)^2+ r^2\left(\frac{d\varphi}{d\tau}\right)^2- \left(1- \frac{\tilde{r}_S}{r}\right)^{r_{dyn}/\tilde{r}_S} \left(\frac{dct}{d\tau}\right)^2= -c^2.\label{c}
\end{align}
With the definitions $u=r^{-1}$ and $'=d/d\varphi$, using the equations (\ref{cb}) and (\ref{ca}), the equation (\ref{c}) reads as follows,
\begin{align}
   -c^2= (1+ \tilde{r}_S u) C_b^2 u'^2+ C_b^2u^2- \frac{C_a^2}{(1- \tilde{r}_S u)^{r_{dyn}/\tilde{r}_S}}.\label{c2}
\end{align}
The relation between the effective and the dynamical radii is given as follows,
\begin{align}
  \frac{\tilde{r}_S- r_{dyn}}{\tilde{r}_S}= \frac{2A}{B}.
\end{align}
Hence, the equation (\ref{c2}) reads for small Schwarzschild radii,
\begin{align}
    C_b^2 u'^2+ C_b^2u^2 (1- \tilde{r}_Su)- C_a^2 (1- \tilde{r}_S u)^{2A/B}= -c^2 (1- \tilde{r}_Su).\label{c5}
\end{align}
After a further derivative in $\varphi$, and considering small effective Schwarzschild radii, the equation (\ref{c5}) leads to
\begin{align}
    u''+ u \left(1- \frac{C_a^2}{C_b^2} A\frac{\tilde{r}_S^2}{r_{dyn}}\right)= \frac{3}{2}\tilde{r}_S u^2+ \frac{\tilde{r}_S}{2C_b^2}\bar{X}c^2,\label{u''Gl}
\end{align}
with the parameter $\bar{X}$ dependent on $C_a$ as follows,
\begin{align}
    \bar{X}=\left[1- \frac{2A}{r_{dyn}}\frac{C_a^2}{c^2}\right].\label{X}
\end{align}
It is clear that for the linear (quasi-Newtonian) approximation, the equation (\ref{u''Gl}) already leads to a trajectory which shows a perihelion shift dependent on the scalar field via $C_a^2 A \tilde{r}_S^2/(C_b^2 r_{dyn})$. For low-energetic systems, however, the Newtonian Kepler orbit appears as first-order solution,
\begin{align}
    u_0= \frac{\tilde{r}_S}{2C_b^2}c^2(1+ \varepsilon \cos\, \varphi).\label{u0}
\end{align}
In the next-order approximation, and only for linear terms in $\varepsilon \varphi$, we have
\begin{align}
    u_1= \frac{\tilde{r}_S}{2C_b^2}c^2\left[1+ \varepsilon \cos\left(1- \frac{3}{4}\frac{\tilde{r}_S c^2}{C_b^2}\right) \varphi\right].\label{u1}
\end{align}
Equations (\ref{u0}) and (\ref{u1}) give the usual value. The perihelion advance for low-energetic systems is then clearly given by
\begin{align}
    \nabla^2 \varphi_{_P}= \frac{6\tilde{M}G_N}{C_b^2}\pi,
\end{align}
which is formally the usual value within GR. It reads as usual for $w=1/6$ such that $\tilde{M}=M_1$. This is the term which appears in (\ref{s1}) as a constraint. At about such pressure, in the equation (\ref{pots}), there is $hr_{dyn}\approx \tilde{r}_S\approx r_S$. For higher values of pressure, effective and dynamical masses are higher than the luminous mass. This
pressure is introduced through the energy--stress tensor as the one of the ideal fluid. It is naturally caused by the gravitational potential and the scalar field, although it is not particular of it. This relation is further elaborated in the next section.

\subsection{Flat rotation curves of galaxies} \label{frc}
\noindent Let us consider the weak fields for galactic ranges. At galactic-bulge ranges, the scalar-field length scale is expected not to be negligible \cite{Cervantes07a,Bezares07-DM}, and importantly shorter ranges would lead to scalar-field masses which would have to have signaled already by means of particle collisions in high-energy experiments. Furthermore, it may be assumed that such length scales are a reason for deviations of dynamics from usual one of GR and thus at least of some Dark Matter dynamics contributions in form of scalar-field components of DM.\\
Let us further consider the tangential velocity of galaxies which is valid for linear solutions \cite{Cervantes09} as given below,
\begin{align}
  v_t= \sqrt{r \frac{d\Phi}{dr}}.\label{rotvel}
\end{align}
It should be valid for linear behaviour. We will use this relation for a system which is not vacuum, however beyond the galactic bulge, as usual within the analysis of rotation curves in relation with universal density profiles \cite{Cervantes07a,NFW}.\\
Now, the Poisson equation in the weak field limit is obtained as follows,
\begin{align}
  \nabla^2 \left(\Phi+ \frac{c^2}{2}\xi\right)= \frac{3\pi G_N}{c^2}(\epsilon+ 3p).\label{Poissonmod}
\end{align}
The scalar field equation yields
\begin{align}
  \nabla^2 \xi- \frac{1}{L^2}\xi= -\frac{2\pi G_N}{c^4}(\epsilon- 3p).\label{Scalarmod}
\end{align}
Phenomenologically, especially rotation velocity of the spiral galaxies is nearly constant (well known as the problem of flat rotation curves) outside the luminous core as if a spherical halo of non-luminous matter with an extension much greater than the galaxy's visible disc surrounded them \cite{Ostriker73} This is usually analysed in the context of universal halo profiles, whereas a halo of non-hadronic, dark matter is assumed \cite{NFW}. In the same spirit, we hence assume now that the rotation velocity is constant and analyse
the necessary conditions following such case. Dark-matter in the sense of non-Newtonian dynamics shall appear as a consequence of the scalar field, related to the dark-matter profile \cite{NFW}.\\
\noindent The form of the gravitational potential which is necessary to give the flat rotation curves is of the following form,
\begin{align}
  \Phi= v_t^2 \text{ln}(r).
\end{align}
The Poisson equation (\ref{Poissonmod}) together with the scalar field equation (\ref{Scalarmod}) leads to
\begin{align}
  \frac{v_t^2}{r^2}+ \frac{c^2}{2L^2}\xi= \frac{4\pi G_N}{c^2}\hat{\epsilon}.\label{Pes}
\end{align}
The equation (\ref{Pes}) defines a density profile having the following form,
\begin{align}
  \hat{\epsilon}=\epsilon+ \frac{3}{2}p=\frac{v_t^2}{4\pi G_N r^2}+ \frac{\xi c^2}{8\pi G_N L^2}.\label{dmp}
\end{align}
It possesses, on the one hand, a contribution $\epsilon$ of matter density in general, and a contribution $p$ of pressure (coming from the inner structure of matter). On the other hand, it has a Newtonian-type energy density and a scalar-field contribution to density distribution as clearly evident from the equation (\ref{dmp}). Hence, we define two energy-density components
as follows,
\begin{alignat}{1}
  \epsilon^*=& \frac{(v_t)^2}{4\pi G_N r^2},\label{es}\\
  \epsilon_\xi=& \frac{\xi c^2}{8\pi G_N L^2}.\label{ex}
\end{alignat}
Both the definitions $\epsilon^*$ and $\epsilon_\xi$ together give the density profile usually called Dark Matter profile $\epsilon_{DM}$ \cite{Cervantes07a,nils-prom}. In these terms, the scalar-field contribution ($\epsilon_\xi$) is expected to act as dark-matter density contribution to the total energy density. The remaining contribution ($\epsilon^*$) is purely
Newtonian and represents the energy density especially of baryons. Furthermore, the scalar field cannot be its own source, which means that it has only usual matter density ($\epsilon^*$) as a source term. For equation (\ref{Scalarmod}), one must have
\begin{align}
  \nabla^2 \xi- \frac{1}{L^2}\xi= -\frac{2\pi G_N}{c^4}\epsilon^*.
\end{align}
The pressure is then given accordingly by the following relation,
\begin{align}
  p= \frac{2}{9}\epsilon_\xi\,.\label{px}
\end{align}
The equation (\ref{px}) shows that the pressure is linearly dependent on the scalar-field density and on the scalar field itself. The scalar-field excitation in such cases is given as follows,
\begin{align}
  c^2\xi= 36 \pi G_N L^2 p\,.\label{xp}
\end{align}
Here, it is clear how scalar-field excitations and pressure are related as a consequence of the baryonic source of scalar fields. The pressure terms (which are first introduced through the ideal fluid) within dark-matter dominance are related to scalar fields and their finiteness as dominant contributions to dark matter.\\
Phenomenologically, there is a relation of about ten to one between hadronic matter ($\epsilon^*$) and dark matter. According to the equations (\ref{es}) and (\ref{ex}), dark matter may then be given by the scalar-field contribution of density. Hence, a relation of the following form is expected,
\begin{align}
  \epsilon_\xi \approx 10\cdot \epsilon^*.
\end{align}
The total energy density $\hat{\epsilon}$ for the dark-matter density profile is in fact given according to equation (\ref{dmp}), and it is interesting that following equation (\ref{px}), the relation between total energy density $\hat{\epsilon}$ and pressure gives an EOS parameter as follows,
\begin{align}
  \hat{w}= \frac{p}{\hat{\epsilon}}\approx \frac{1}{5}\,.\label{pet5}
\end{align}
For large, galactic scales, hence, $\hat{w}$ is given by the dark-matter contribution which comes from the scalar field. Furthermore, for vanishing contributions of the scalar field, $p/\hat{\epsilon}_\xi$ tends to zero, and for $\epsilon^*=0$, i.e. for a complete dominance of the scalar-field excitation, the total EOS parameter reads exactly 1/5. Astonishingly, this value which is necessary within dark-matter phenomenology of flat rotation curves is equivalent to the EOS parameter $w$ within the context of solar-relativistic effects. It appears for matter density given by usual matter $\epsilon^*$ with Newtonian dynamics and for such matter as only source of the scalar field. Apparently, an EOS parameter of $1/6<w<1/5$ is a weak-field constraint not only for solar-relativistic effects and energy density of gravitation but also within dark matter phenomenology derived from the present induced gravity model with a Higgs potential. The behaviour of the contributions of pressure, however, differs for both cases. We will now
investigate the behaviour of density components for galactic dynamics.\\
After parametrising distance by a length scale $a$ of the spherical system (a length related to the distance at which galaxies possess flat rotation curves), in the interval between $r=0$ and $r=r_H$ with $r_H$ as halo radius with $r_H>L$ and $r_H>a$ \cite{Cervantes07a, nils-prom}, the solution of the scalar field yields
\begin{align}
  \xi=\frac{1}{2r_a}\frac{v_t^2}{c^2}\left[e^{-\frac{r_a}{l_a}} \, \text{Shi} \left(\frac{r_a}{l_a}\right)- \sinh\left(\frac{r_a}{l_a}\right)\text{Ei} \left(-\frac{r_a}{l_a}\right)\right],\label{xiw}
\end{align}
whereas $r/a=r_a$ and $L/a=l_a$, Shi$(x)$ is the hyperbolic sine integral function (i.e. SinushIntegral$(x)$) and Ei$(x)$ is the exponential integral function. For the dark-matter profile (total density distribution), we obtain
\begin{align}
  \hat{\epsilon}= \frac{(v_t c)^2}{4\pi G_N a^2}\left\{\frac{1}{r_a^2}+ \frac{1}{4l_a^2r_a}\left[e^{-\frac{r_a}{l_a}} \, \text{Shi} \left(\frac{r_a}{l_a}\right)- \sinh\left(\frac{r_a}{l_a}\right)\text{Ei} \left(-\frac{r_a}{l_a}\right)\right]\right\}. \label{hate}
\end{align}
It gives the halo structure in a way analogous to the Navarro--Frenk--White (NFW) profile with scale radii $r_s=a$ \cite{NFW}. Such  models are used in search of the universal halo densities in the context of flat rotation curves.\\
\begin{figure}[h] \centering
\includegraphics*[width=14.5cm]{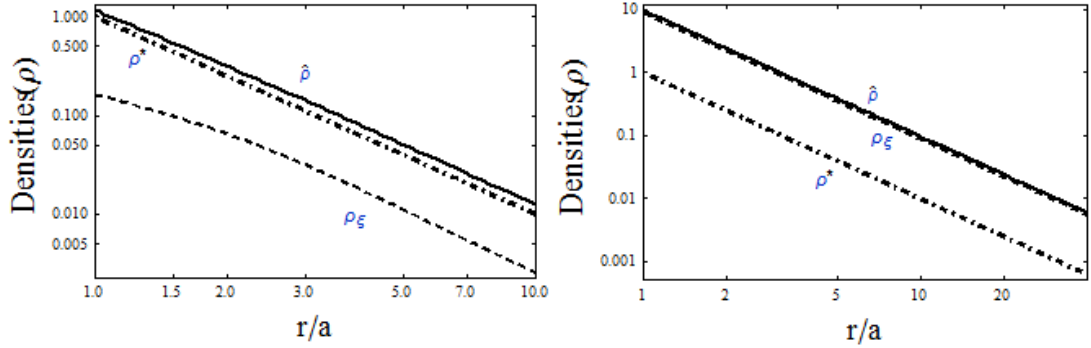}
\caption{Evolution of density distributions normalised to $v_tc^2/(4\pi G_Na^2)$ for $l_a=1$ (left panel) and $l_a=1/35$ (right panel). Scalar field ($\epsilon_\xi$) dominance for shorter distances and baryonic ($\epsilon^*$) dominance for limits of large scales.}
\label{fig4}
\end{figure}
The scale radius $a$ is of the order of magnitude of a galactic core $R_1$ (i.e. the luminous-disc radius of galaxies). For higher values of $L$ in relation to $a$, baryonic (usual) density ($\epsilon^*$) dominates. However, for low values of the scale factor $L$ in relation to $a$, the scalar-field density ($\epsilon_\xi$) dominates. This may be seen in Fig. \ref{fig4}.
Further, in the left panel of Fig. \ref{fig4}, a relation between the total density profile and baryonic density (i.e. a ratio of ten to one) is visible.\\
Let us now define the ratio of the density parameters by using equations (\ref{es}) and (\ref{hate}),
\begin{align}
    \Delta\equiv \hat{\epsilon}/\epsilon^*= 1+ \frac{r_a}{l_a^2}\left[e^{-r_a/l_a}\text{Shi}(r_a/l_a)- \text{sinh}(r_a/l_a)\text{Ei}(-r_a/l_a)\right].
\end{align}
\begin{figure}[h] \centering
\includegraphics*[width=14.5cm]{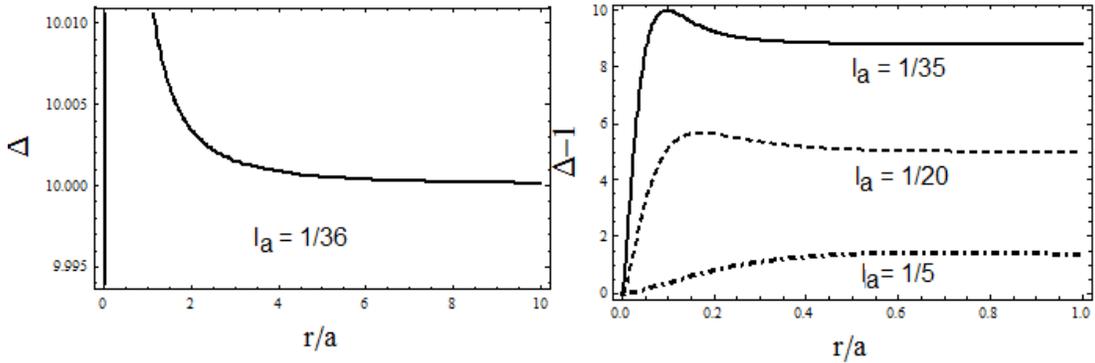}
\caption{Density ratios: Dark matter dominance for $l_a=1/36$ (left panel) and non-Newtonian behaviour (right panel) for $l_a=1/5$, $l_a=1/20$ and $l_a=1/35$.}
\label{fig5}
\end{figure}
\noindent The density ratio gives non-baryonic behaviour ($\Delta-1\neq 0$), and it shows three special cases. At lower scales (as shown in the right panel of Fig. \ref{fig5}), a linearly growing function with relatively high slope, at high scales a constant value, and an intermediate phase with a maximum (viz Fig. \ref{fig5}). For all length scales $l_a$, the nonbaryonic behaviour
$\Delta- 1$ is negligible at shorter ranges, even though scalar-field densities do dominate. Hence, the dominant scalar-field contribution of density acts as a baryonic contribution for shorter distances (even $r_a>1$).\\
For $L/a\approx 35$ (see left panel of Fig. \ref{fig5}), $\Delta\approx 10$ (i.e. long-range dynamics are as if there were 10 times the baryonic density). There is scalar-field density ($\epsilon_\xi$) dominance at distances of galactic bars, and the relation $\hat{w}=1/5$ is valid and non-Newtonian behaviour of the scalar field is thus dominant for flattening dynamics of
galaxies.\\

\section{Conclusions and Future Directions}
In this article, we have investigated the relation between the scalar-field excitations of induced gravity with a Higgs potential and the obligatory presence of finite pressure terms in energy density of gravitation, along with their appearance in linear solutions for solar-relativistic effects
such as perihelion advance, and for flat rotation curves leading to Dark Matter phenomenology with scalar-field density components of the dark-matter profile. The important conclusions drawn from this study are summarised below:
\begin{itemize}
  \item [(i)] An energy density of gravitation following Maxwell-like equations may differ with its analogue of GR. Gravitational energy within induced gravity and GR are identical for $\hat{q}=0$ which denotes the coupling of the scalar field to the matter Lagrangian as well as a decoupled Higgs field, and for $\hat{q}=1$ (i.e the absence of the coupling of the scalar field with the matter Lagrangian) they are the same only with the constraint on the EOS parameter as  $1/6<w<1/5$.
  \item [(ii)] Finite values of pressure of the ideal fluid within vacuum solutions are expected from the nature of scalar-field excitations. This further leads to the notion of the dynamic and bare (luminous) masses in this model. The value of the dynamical mass is observed greater than that of the luminous mass, and the present formulation is thus useful to describe the signatures of unseen matter in nature.
  \item [(iii)] Perihelion shift is found the same as within GR for low-energetic systems with values of pressure as constrained by energy in view of the EOS parameter as $w=1/5$ as in (i). There is, further, more complicated dynamics for highly energetic systems with high coefficients of $C_a$ per $c$ and $C_b$. Scalar fields are essential for low-range dynamics although they have Newtonian behaviour at such range.
  \item [(iv)] Flat rotation curves of galaxies lead to dark-matter profiles with baryonic and scalar-field components of density. Scalar-field densities are strongly related to pressure terms in this model and they seem to provide a viable explanation of the Dark Matter contents of our Universe.
  \item [(v)] Dark Matter dominance leads to pressures related to an EOS parameter of total energy of the same value as for weak fields in solar-relativistic ranges.
  \item [(vi)] The non-Newtonian behaviour of density appears as distances grow, according to (iii). Such non-Newtonian behaviour of scalar-field excitations leads to flat rotation curves. The contribution due to the scalar field in the energy density in fact acts as the dark-matter profile in view of the total energy density of the system.
\end{itemize}
However, the estimation of the shift of intermediate behaviour of scalar fields, i.e. the relation between scalar fields and galactic centres, is still unclear, and it would be a quite interesting problem to investigate. Further, this might also be valuable in relation to quintessential properties of scalar fields for galaxies within exact solutions, leading to Reissner--Nordstr\"{o}m
behaviour. Moreover, the cosmological implications of induced gravity with Higgs potentials in terms of the quintessence and dark matter along with the primeval dynamics would be another important task to study in greater detail. We intend to report on such issues in our forthcoming communication \cite{nils2010}.

\section*{Acknowledgments}
\small{One of the authors (HN) is thankful to the University Grants Commission, New Delhi, India for financial support under the UGC--Dr. D. S. Kothari post-doctoral fellowship programme. The work of HN was also supported by the German Academic Exchange Service (DAAD), Bonn, Germany in terms of a study visit to the Department of Physics, University of Konstanz, Germany. HN and NB would also like to thank Prof. F. Steiner and other members of the Cosmology and Quantum Gravitation Group of the Institute of Theoretical Physics, University of Ulm, as well as the Graduate School for Mathematical Analysis of Evolution, Information and Complexity for their kind support during the course of this work.}\\


\begin{center}
\section*{Appendix I}
\end{center}
\renewcommand{\theequation}{A-\arabic{equation}}
Following the Ricci identities, the Ricci tensor may be written as follows,
\begin{align}
  -R^\sigma\,_\mu u_\sigma= (u^\sigma\,_{;\mu}- u^\alpha\,_{;\alpha} \delta_\mu\,^\sigma)_{;\sigma} =H^\sigma\,_{\mu;\sigma},\label{a1}
\end{align}
where $H_{\mu \nu}=u_{\nu;\mu}- u^\alpha\,_{;\alpha}\,  g_{\mu \nu}$ is an antisymmetric tensor. The (antisymmetric) field-stress tensor $\tilde{F}_{\mu \nu}$ in terms of $H_{\mu \nu}$  as defined in (\ref{a1}) follows as below,
\begin{align}
    \tilde{F}_{\mu \nu}=H_{\nu \mu}- H_{\mu \nu}= u_{\nu; \mu}- u_{\mu; \nu}.
\end{align}
Further, a symmetric tensor $Q_{\mu \nu}$ is defined as below,
\begin{alignat}{1}
    Q_{\mu \nu} \equiv  H_{\nu \mu}+ H_{\nu\mu}=\,& u_{\mu; \nu}+ u_{\nu;\mu}- 2u^\alpha\,_{;\alpha}g_{\mu \nu}\label{Q}.
\end{alignat}
The tensor $Q_{\mu \nu}$ (\ref{Q}) is related to the Ricci tensor and the generalised field-strength tensor (\ref{fieldst}) as given below,
\begin{align}
    -R^\lambda\,_\mu u_\lambda= \frac{1}{2}\tilde{F}_\mu\,^\lambda\,_{;\lambda}+ \frac{1}{2}Q_\mu\,^\lambda\,_{;\lambda}\,.\label{rel}
\end{align}
However, the divergence of $Q_\mu\,^\lambda$ in (\ref{rel}) is given by
\begin{align}
  Q_\mu\,^\lambda\,_{;\lambda}= 2u^\lambda\,_{;\mu;\lambda}- 2u^\lambda\,_{;\lambda;\mu}. \label{qmu}
\end{align}
Given the unitarity of the metric, the equation (\ref{qmu}) then leads to
\begin{align}
  Q_\mu,^\lambda\,_{;\lambda}u^\mu= 4(u^\lambda\,_{;\mu} u^\mu)_{;\lambda}- 4u_\mu\,^{\lambda} u^\mu\,_{;\lambda}- 2u^\lambda\,_{;\mu;\lambda}u^\mu- 2u^\lambda\,_{\lambda;\mu}u^\mu. \label{qmulam}
\end{align}
For the static field in relation to the observer, the equation (\ref{qmulam}) gets further simplified as follows,
\begin{align}
  Q_\mu\,^\lambda\,_{;\lambda}u^\mu= -4u_\mu^{;\lambda} u^\mu\,_{;\lambda}- 2u^\lambda\,_{;\mu;\lambda}u^\mu.\label{Qs}
\end{align}

\newpage
\begin{center}
\section*{Appendix II}
\end{center}
In static linear approximation for potentials $\nu$ and $\lambda$ along with $\bar{\xi}= 1+(3\xi/2)/(1+ \xi)$, the scalar-field and Einstein equations in spherical symmetry for a perfect fluid read
\begin{alignat}{1}
  & \nabla^2 \xi- \frac{1}{L^2}\xi= -\hat{q}\, \frac{8\pi G}{3\,c^4}\, (\epsilon- 3p),\label{sfa}\\
  &\nabla^2\nu + \frac{\bar{\xi}}{L^2}= \frac{16\pi \tilde{G}}{c^4}\left[\left(\epsilon- \frac{1}{2}\left(1- \frac{\hat{q}}{3}\right)(\epsilon- 3p)\right)\right],\label{nfa}\\
  & \nu''- \frac{2}{r}\lambda'+ \frac{\bar{\xi}}{L^2}= \frac{16\pi \tilde{G}}{c^4}\left[p- \frac{1}{2}\left(1- \frac{\hat{q}}{3}\right)(\epsilon- 3p)\right].
\end{alignat}
Using equations (\ref{sfa}) and (\ref{nfa}), the equation for a potential defined with $\Psi= \frac{c^2}{2}(\nu+ \xi)$ reads as follows,
\begin{align}
  \nabla^2 \Psi (1+ \xi)= \frac{4\pi G}{c^4}\left[\epsilon+ 3p- \frac{\hat{q}}{2}\xi(\epsilon-
3p)\right]- \frac{c^2}{4}\xi\nabla^2 \xi. \label{poissonappin}
\end{align}
The equation (\ref{poissonappin}) reduces to the usual Poisson equation for finite density and pressure terms in case of linear behaviour of $\xi$,
\begin{align}
    \nabla^2 \Psi= \frac{4\pi G}{c^4}(\epsilon+ 3p).
\end{align}
In fact, a contribution of the scalar field is added to the usual gravitational potential $\Phi= \nu c^2/2$.\\


\begin{thebibliography}{99}
\bibitem{Sami} M. Sami, \emph{Curr. Sci. {\bf 97}} (2009) 887 (arXiv:0904.3445 [hep-th]).
\bibitem{Sahni} V. Sahni, astro-ph/0403324v3.
\bibitem{Copeland} E.J. Copeland, M. Sami and S. Tsujikawa, \emph{Int. J. Mod Phys. {\bf D15}} (2006) 1753.
\bibitem{Sami1} M. Sami, \emph{Lect. Notes Phys. {\bf 720}} (2007) 219.
\bibitem{Xu} L. Xu and J. Lu, \emph{Eur. Phys. J. {\bf C60}} (2009) 135.
\bibitem{Neupane} I.P. Neupane and C. Scherer, \emph{JCAP {\bf 05}} (2008) 009
(arXiv:0712.2468 [astro-ph]).
\bibitem{Das} S. Das and N. Banerjee, \emph{Phys. Rev. {\bf D78}}
  (2008) 043512.\\
N. Banerjee and D. Pav\'on, \emph{Phys. Lett. {\bf B647}} (2007) 477.
\bibitem{Feng}  Ch.-J. Feng, arXiv:0806.0673.\\
    E. Elizalde, S. Nojiri, S. Odintsov, D. S\'aez - G\'omez and V. Faraoni, hep-th/0803.1311.
\bibitem{Brun}  T. Brunier, V. K. Onemli, R.P. Woodard, \emph{Class. Quantum Grav. {\bf 22}} (2005) 59.\\
    O. Bertolami, P.J. Martins, \emph{Phys. Rev. {\bf D61}} (2000) 064007.
\bibitem{Ratra} P.J.E. Peebles and B. Ratra, \emph{Rev. Mod. Phys. {\bf 75}} (2003) 559.
\bibitem{Padmanabhan} T. Padmanabhan, \emph{Phys. Rept. {\bf 380}} (2003) 235
(arXiv:hep-th/0212290v2).
\bibitem{Kinney} W.H. Kinney, arXiv:0902.1529v2 [astro-ph.CO].
\bibitem{Kozyrev08} S.M. Kozyrev, arXiv:0808.3322 [gr-qc].
\bibitem{Rod09} M.A. Rodr\'{i}guez-Meza, arXiv:0907.2898 [astro-ph.CO].
\bibitem{Bijay} B.K. Sahoo and L.P. Singh, \emph{Mod. Phys. Lett. {\bf A17}} (2002) 2409;
\emph{ibid.} {\bf A18} (2003) 2725.
\bibitem{Campo} S. del Campo, R. Herrera and P. Labrana, \emph{JCAP {\bf 711}} (2007) 30.
\bibitem{Bibeka} B. Nayak and L. P. Singh, Mod. Phys. Lett. A {\bf 24} (2009) 1785 (arXiv:0803.2930 [gr-qc]).
\bibitem{Jae-Weon Lee1} Jae-Weon Lee, Phys.Rev. D {\bf 53} (1996) 2236 (hep-ph/9507385).
\bibitem{Jae-Weon Lee2}  Jae-Weon Lee, arXiv:0801.1442v4 [astro-ph].
\bibitem{Higgs} P.W. Higgs, \emph{Phys. Rev. Lett. {\bf 12}} (1964) 132.
\bibitem{Brans} C.H. Brans, gr-qc/0506063. Contributions to the Cuba Workshop, ``Santa Clara 2004. I International Workshop on Gravitation an Cosmology''.
    \bibitem{Dehnen1} H. Dehnen and H. Frommert, \emph{Int. J. Theor. Phys. {\bf 29}(4)} (1990) 361; \emph{ibid. {\bf {29}}(6)} (1990) 537; \emph{ibid.
    {\bf 30}} (1991) 985.
\bibitem{Dehnen2} H. Dehnen, H. Frommert and F. Ghaboussi, \emph{Int. J. Theor. Phys. {\bf 31}} (1992) 109.
\bibitem{Dehnen3} H. Dehnen and H. Frommert; \emph{Int. J. Theor. Phys. {\bf 32}(7)} (1993) 1135.
\bibitem{Zee} A. Zee, \emph{Phys. Rev. Lett. {\bf 42 (7)}} (1979) 417.
\bibitem{Cervantes95a} J.L. Cervantes-Cota and H. Dehnen, \emph{Phys. Rev. {\bf D51}} (1995) 395.\\
        J.L. Cervantes-Cota and H. Dehnen, \emph{Nucl. Phys. {\bf B442}} (1995) 391.
\bibitem{Bezares07-DM} N.M. Bezares-Roder and H. Dehnen, \emph{Gen. Rel. Grav. {\bf 39}}(2007) 1259 (arXiv:0801.4842[gr-qc]).
\bibitem{Cervantes07a} J.L. Cervantes-Cota, M.A. Rodr\'{\i}guez-Meza, D. N\'u\~nez, \emph{J. Phys. Conf. Ser. {\bf 91}} (2007) 012007.
\bibitem{Bezares06H} N.M. Bezares-Roder and H. Nandan, \emph{Indian J. Phys. {\bf 82}} (2008) 69 (arXiv:hep-ph/0603168).
\bibitem{nils-prom} N.M. Bezares-Roder and F. Steiner in  \emph{Mathematical Analysis of Evolution, Information and Complexity}, eds. Arendt and Schleich (Wiley-VCH, Berlin, 2009).
\bibitem{Bij94} J.J. van der Bij, \emph{Acta Phys. Pol. {\bf B25}} ({1994}) 827.
    arXiv:9319964.
\bibitem{Bezares09-RN} H. Nandan, N.M. Bezares-Roder and H. Dehnen, Class. Quantum Grav. {\bf 27} (2010) 245003 (arXiv:0912.4036[gr-qc]).
\bibitem{Peskin}  M.E. Peskin and D.V. Schroeder, \emph{An Introduction to Quantum Field Theory} (Addison-Wesley, New York, 1995).
\bibitem{Wyman} M. Wyman, \emph{Phs. Rev. {\bf D24}} ({1981}) 839;\\
    P. Baekler, E.W. Mielke, R. Hecht and F.W. Hehl, \emph{Nucl. Phys. {\bf B288}}
    ({1987}) 800;\\
B.C. Xanthopoulos and T. Zannias, \emph{Phys. Rev. {\bf D40}} ({1989}) 2564.
\bibitem{Schmoltzi} K. Schmoltzi and Th. Sch\"uker, \emph{Phys. Letts. {\bf 161}}
({1991}) 212.
\bibitem{Jetzer} P. Jetzer and D. Scialom, \emph{Phys. Letts. {\bf A169}} ({1992}) 12.
\bibitem{Hardell} A. Hardell and H. Dehnen, \emph{Gen. Rel. and Grav. {\bf 25}(11)} ({1993}) 1165.
\bibitem{Vir} K.S. Virbhadra, S. Jhingan and P.S. Joshi, \emph{Int. J. Model. Phys. {\bf D6}} ({1997}) 357;
 K.S. Virbhadra, \emph{Phys. Rev. {\bf D60}} ({1999}) 104041;
 K.S. Virbhadra and G.F.R. Ellis, \emph{Phys. Rev. {\bf D62}} ({2000}) 084003;
K.S. Virbhadra and G.F.R. Ellis, \emph{Phys. Rev. {\bf D65}}, ({2002}) 103004.
\bibitem{Penrose98} R. Penrose in ``Black Holes and Relativistic Stars'', edited by R.W. Wald, The University of Chicago Press, Chicago (1998), p. 103.
\bibitem{Ori} A. Ori and T. Piran, \emph{Phys. Rev. Lett. {\bf 59}} ({1998}) 2137;
    A. Ori and T. Piran, \emph{Phys. Rev. {\bf D42}} ({1990}) 1068.
\bibitem{Joshi1} P.S. Joshi and I.H. Dwivedi, \emph{Phys. Rev. {\bf D47}} ({1993})
    5357.
\bibitem{JoshiB} P.S. Joshi, ``Gravitational Collapse and Spacetime Singularities'', Cambridge University Press(2007).
\bibitem{Jhingan} S. Jhingan and G. Magli, \emph{Phys. Rev. {\bf D61}} ({2007})
    124006.
\bibitem{Shapiro91} S.L. Shapiro and S.A. Teukolsky, \emph{Phys. Rev. Lett. {\bf 66}} ({1991}) 994;
    S.L. Shapiro and S.A. Teukolsky, \emph{Phys. Rev. {\bf 45}} ({1991}) 2006.
\bibitem{Zade} S.S. Zade, D.K. Pati, N.P. Mulkalwar, \emph{Chin. Phys. Lett. {\bf 25}(5)} ({2008}) 968.
\bibitem{Bronnikov} K.A. Bronnikov  and  M.S. Chernakova,
\emph{Grav. Cosmol. {\bf 13}} ({2007}) 51;\\
    H. Alavirad and  P. Ameri, arXiv:0901.0262.
\bibitem{Sheykhi} A. Sheykhi and H. Alavirad, \emph{Int. J. Mod. Phys. {\bf D18}}
    ({2009}) 1773.
\bibitem{Kozyrev} S.M.Kozyrev and S.V. Sushkov, arXiv:0812.5010.
\bibitem{Nandi}  K.K. Nandi, B. Bhattacharjee, S.M.K. Alam, J. Evans,
\emph{Phys. Rev. {\bf D57}} ({1998}) 823.
\bibitem{moffat} J.M. Moffat, gr-qc/0702070.
\bibitem{Chakrabarti} S.K. Chakrabarti and P.S. Joshi, \emph{Int. J. Mod. Phys. {\bf D3}} ({1994}) 647.
\bibitem{Joshi2} P.S. Joshi, \emph{Gen. Rel. Grav. {\bf 30}(11)} ({1998}) 1563.
\bibitem{Harko03} T. Harko, \emph{Phys. Rev. {\bf D68}} ({2003}) 064005.
\bibitem{Haw72} S.W. Hawking, \emph{Commun. Math. Phys. {\bf 25}} ({1972}) 167.
\bibitem{Cam93} M. Campanelli and C.O. Lousto, \emph{Int. J. Mod. Phys. D {\bf 2}} ({1993})
451. 
\bibitem{Bhadra} A. Bhadra and K. Sarkar, \emph{Gen. Rel. Grav. {\bf 37} }({2005})
    2189.
\bibitem{Bezares07-BH} N.M. Bezares-Roder, H. Nandan and H. Dehnen, \emph{Int. J. Theor. Phys. {\bf 46}(10)} ({2007}) 2429 (arXiv:gr-qc/0609125).
\bibitem{Avelino} P.P. Avelino, A.J.S. Hamilton and C.A.R. Herdeiro, \emph{Phys. Rev. {\bf D79}} ({2009}) 124045.
\bibitem{Lopez} M. Bouhmadi-L\'{o}pez and D. Wands, \emph{Phys. Rev. {\bf D71}}
    ({2005}) 024010.
\bibitem{Narayana} N. Banerjee, S. Sen and N. Dadhich, \emph{Mod. Phys. Lett.  {\bf A16}} ({2001}) 1223.
\bibitem{Rei16} H. Reissner, \emph{Ann. der Physik {\bf 50}} ({1916}) 106.
\bibitem{vd2009}V. Dzhunushaliev, K. Myrzakulov and R. Myrzakulov, aXiv:0907.5265 [gr-rc].
\bibitem{Cervantes07} J.L. Cervantes-Cota and J. Rosas \'Avila, \emph{Rev. Mex. F\'{\i}s. {\bf S53}} ({2007}) 137.
\bibitem{Adelberger} E.G. Adelberger \emph{et al.}, \emph{Phys. Rev. Lett. {\bf 98}} ({2007}) 131104.
\bibitem{Dehnen64a} H. Dehnen, \emph{Ann. d. Phys. {\bf 7}(10)} ({1964}) 101.
\bibitem{Dehnen64b} H. Dehnen, \emph{Zs. f. Phys. {\bf A179}(1)} ({1964}) 96.
\bibitem{Wetterich02} C. Wetterich, Phys. Rev. {\bf D65} ({2002}) 123512
    (arXiv:hep-ph/0108266).
\bibitem{Wetterich03} C. Wetterich, arXiv:hep-ph/0203266.
\bibitem{Cervantes09} J.L. Cervantes-Cota, M.A. Rodr\'{i}guez, D. N\'{u}\~{n}ez,
    \emph{Phys. Rev. {\bf D79}} ({2009}) 064011.
\bibitem{Doran02} M. Doran and C. Wetterich, arXiv:astro-ph/0205267.
\bibitem{Wetterich01} C. Wetterich, arXiv: astro-ph/0108411.
\bibitem{Kroupa} P. Kroupa \emph{et al.}, \emph{Astronomy \& Astrophysics}, (to appear), arXiv:1006.1647v2[astro-ph.CO].
\bibitem{NFW} J.F. Navarro, C.S. Frenk and S.D.M. White, \emph{Astrophys. J. {\bf 462}} ({1996}) 563.
\bibitem{Ostriker73} J.P. Ostriker and P.J.E. Peebles, \emph{Astrophys. J. {\bf 186}} ({1973}) 476.
\bibitem{nils2010} N.M. Bezares-Roder, Hemwati Nandan and U.D. Goswami (2010) in preparation.
\end{thebibliography}
\end{document}